\documentclass[journal]{IEEEtran}

\ifCLASSINFOpdf
\else
\fi
\usepackage{graphicx}
\usepackage{color}
\usepackage{subcaption}
\usepackage{textcomp}
\usepackage{verbatim}
\usepackage[numbers,sort]{natbib}
\begin{document}
%
\title{Dense Deformation Network for High Resolution tissue-cleared Image Registration}
%
%
%

\author{
	\IEEEauthorblockN{Abdullah Nazib, Clinton Fookes and Dimitri Perrin\IEEEauthorrefmark{1}}\\
	\vspace{5mm}
	School of Electrical Engineering \& Computer Science,\\Queensland University of Technology, Brisbane, Australia.\\
	\IEEEauthorblockA{\IEEEauthorrefmark{1}Corresponding author; dimitri.perrin@qut.edu.au}
}
%


\maketitle

\begin{abstract}
The recent application of deep learning in various areas of medical image analysis has brought excellent performance gains. In particular, technologies based on deep learning in medical image registration can outperform traditional optimisation-based registration algorithms both in registration time and accuracy. However, the U-net based architectures used in most of the image registration frameworks downscale the data, which removes global information and affects the deformation. In this paper, we present a densely connected convolutional architecture for deformable image registration. Our proposed dense network downsizes data only in one stage and have dense connections instead of the skip connections in U-net architecture. The training of the network is unsupervised and does not require ground-truth deformation or any synthetic deformation as a label. The proposed architecture is trained and tested on two different versions of tissue-cleared data, at 10\% and 25\% resolution of the original single-cell-resolution dataset. We demonstrate comparable registration performance to state-of-the-art registration methods and superior performance to the deep-learning based VoxelMorph method in terms of accuracy and increased resolution handling ability.
In both resolutions, the proposed DenseDeformation network outperforms VoxelMorph in registration accuracy. 
Importantly, it can register brains in one minute where conventional methods can take hours at 25\% resolution.
\end{abstract}


\IEEEpeerreviewmaketitle
\section{Introduction}
\IEEEPARstart{}{Traditionally} image registration problem is addressed as an optimisation problem that maximises similarity between images until registration parameters that achieve satisfactory performance are obtained. For a deformable registration problem, a large number of parameters are required and computing them is computationally expensive. The number of parameters and their computation expense increases with the resolution and dimension of the images. For high-resolution images, a deformable image registration algorithm takes a very long runtime, which reduces its inapplicability. Recently, a bio-chemical process named tissue clearing has emerged that can remove light-obstructing elements from soft tissues and enable biologists to take 3D images of entire organs at high resolution. A number of tissue clearing methods have been developed like BABB~\cite{Dodt2007}, Scale~\cite{Hama2011}, SeeDB~\cite{Ke2013}, CLARITY~\cite{Chung2013}  and iDISCO~\cite{Renier2014}. In this paper we used data from a tissue-clearing method named CUBIC~\cite{Susaki2014}. The 3D images obtained from tissue clearing and light-sheet fluorescence microscopy (LSFM) are very large. Compared to MRI data, a brain sample from a tissue-clearing protocol is 1000 times larger and requires gigabytes of space. In terms of resolution, tissue-cleared volumes are in micrometre scale $(6.45\times6.45\times10\mu m^3)$ whereas MRI volumes typically are in millimetre scale $(0.86\times0.86\times1.5mm^3)$. Therefore, the registration of these images through conventional registration methods is computationally very expensive~\cite{Nazib2018}. While they have been used in recent studies involving tissue clearing (see e.g.~\cite{Tatsuki2016,NIWA20182231}), they are not practical as the number of brains to analyse increases.\\

The recent application of deep learning in image registration provides a promising way to address this issue~\cite{Balakrishnan2018,Eppenhof2019,Cao2018}. The computational efficiency of these methods is extremely high. They take only a few seconds to register images, and have satisfactory registration accuracy. However, training a deep-learning registration method requires large amounts of data. In our context, having large amount of tissue-cleared data (similar to~\cite{Balakrishnan2018}) is not practical given how recent the clearing techniques are. Moreover, training of a deep-learning based registration method on high resolution tissue-cleared data requires huge storage and computational memory (RAM) that is not always feasible. In addition to the issues mentioned, some methods~\cite{Balakrishnan2018} require retraining for new reference image and are not applicable for arbitrary image registration like iterative registration methods.
In summary, registering high-resolution tissue-cleared data using deep-learning requires following criterion to be met:

\begin{itemize}
    \item The learning framework should be trainable with small amount of data.
    \item The learning framework should have the ability to take high resolution images with limited computational resources.
    \item It should be scalable in order to handle higher resolution.
    \item It should not be dependent on a specific reference image, so that any arbitrary image pair can be registered without further training.   
\end{itemize}

To address these issues, we propose a patch-based densely connected image registration network. The proposed network generates deformation in two steps between two dense blocks. These blocks are densely connected convolutional layers able to capture complex features from the tissue-cleared data. Thus, the generated deformation fields can capture small deformations in the cellular structures and give better results than auto-encoder based learning frameworks. In an auto-encoder based architecture, the successive down-scaling removes the global information from the training patches. The proposed network generates a deformation field from the dense block without down-scaling and hence is able to capture global deformation more accurately.
The tissue-cleared images contain cellular information and the intensities are not continuous but discrete. The difference of cellular structure between two brains is also small. Hence, the deformation in the tissue-cleared brain images are small and smooth. To capture small deformations, the second dense block extracts convolutional features from the down-sampled feature bank obtained from the first dense block. Hence, a dense deformation field is obtained, and the network is named as the Dense Deformation Network (DDN). The proposed network is scalable in resolution and trainable with limited computational resource.

\section{Related Works}
Deep learning has been extensively used and investigated in medical image analysis, such as for medical image segmentation. Recent work has shown that it may also be applicable to image registration. There are two prevalent approaches to use deep learning based image registration~\cite{Lai}, and these are outlined in the following subsections. 

\subsection{Deep Learning for Similarity Measure}
In this approach, deep networks are used to provide a similarity measure~\cite{Cheng2016,Simonovsky2016,Wu2016}. In~\cite{Cheng2016} a DNN was first used for multi-modal image registration. Two stacked auto-encoder (SAE) networks and a prediction layer are used to measure similarity between moving and fixed image patches in an unsupervised manner. The prediction layers determine the similarity between the image patches. A similar strategy is proposed in~\cite{Simonovsky2016}, where a CNN is integrated into an optimisation based registration framework. After re-sampling the moving image by the registration process, the re-sampled moving image and fixed image patches are fed into the CNN architecture. The CNN then outputs a dissimilarity map. The derivative of the dissimilarity map is then directly used for parameter optimisation. In~\cite{Wu2016} a deep auto-encoder based registration framework scalable to different modalities is proposed. The encoder learns the low dimensional representation of high dimensional 3D patches and outputs only 1D representation vectors. The decoder reconstructs 3D patches again from the 1D representation vectors. A simple feature based image registration framework named HAMMER is then used for registration using low dimensional representation vectors from the encoder.

\subsection{Deep Learning for Parameter Estimation}
This is a different strategy to use deep learning in image registration. Here, deep learning is directly employed to estimate transformation parameters.  Miao \emph{et al.} first proposed a CNN regression parameter estimator for 2D-3D registration~\cite{Miao2016}. In their method, the rigid transformation parameters are partitioned. A set of CNN networks are used to predict and update the transformation parameters taking DRR (Digitally Reconstructed Radiograph) and X-ray ROIs€™ as input. This approach shows significant performance improvement in several datasets. To capture deformation by deep learning,~\cite{Yang2016} considers LDDMM (Large Deformation Diffeomorphic Metric Mapping) as their baseline. A CNN-based deep architecture is used to predict the momentum maps. This method provides significant performance boost (registration speed) for both 2D (around $1600\times$) and for 3D (around $66\times$) registration. Despite its accuracy, this method requires pre-registration of the training dataset to generate deformation fields on which training of the CNN network is dependent. \\

To make the registration process independent from other baseline methods, Sokooti \emph{et al.}~\cite{Sokooti2017} proposed to train their registration network with randomly generated deformation fields as labels. Using random deformation as labelled data does not guarantee the invertibility and smoothness of the deformation fields. In~\cite{Rohe2017}, the authors proposed segmentation based ground-truth/label data generation, where a segmented region of interest from each training image is registered to a template image giving a set of correspondence points. Deformation fields are then generated by interpolating these correspondence points in a grid. Thus, this method can generate deformation fields between any image pairs in the training data. Still, its dependence on an optimisation algorithm makes it vulnerable to the limitations of the optimisation algorithm used.\\ 

To make learning-based registration algorithm-independent,~\cite{Li2018b} used a fully convolutional structure that directly estimates spatial deformation between the fixed and moving image by optimising similarity and regularisation loss. The network provides loss in different spatial resolutions and combines them to optimise with back-propagation. The training strategy is similar to a conventional registration approach and is independent of the ground-truth deformation label. The first learning-based registration method that is independent and also achieved state-of-the-art performance in 8 different datasets is proposed in~\cite{Balakrishnan2018}. This method combines a U-net FCN architecture with a spatial transformation network (STN) for interpolation, with a loss function based on the cross-correlation between the fixed and interpolated moving images. Rigorous training and testing on eight different MRI datasets with around 8000 images make this voxelmorph (VM) method a viable competitor to conventional approaches. In this paper, we used voxelmorph as one of the baseline methods for comparison.

\section{Training Data Generation}
In this work, the proposed network is trained and tested using tissue-cleared images of the Arc-dVenus mouse brain. The data acquisition process and pre-processing steps have been previously explained~\cite{Susaki2015, Nazib2018}.  
The following is a brief discussion of the training data generation process.      

\subsection{Training Data Generation}
To evaluate the performance of our proposed method and compare it with baseline methods, we chose 10\% and 25\% resolution images of CUBIC brains for training. The spatial dimension at these two resolutions are $256\times216\times68$ and $640\times540\times 169$ respectively. We did not use samples at their original resolution, as existing tools cannot efficiently handle such files~\cite{Nazib2018}.
We have 20 brains for training and 3 brains for testing (in both resolutions). To train the network with these 20 brains we have $20\times19$ pairs of source and target images. Before extracting the training patches, we select brain ``003'' from the test dataset as the reference to all brains and perform affine registration to bring all brains into the same rigid space. Intensity normalisation (between 0 to 1) is applied to all data before extracting the patches.\\

To extract training patches, each source and target images are selected randomly out of 20 brains. From each pair, 2500 patches are randomly picked. To ensure patches are informative and discriminative, we apply canny edge detection on both the source and target images with upper threshold 0.5 and lower threshold 0.02 to obtain binary images. If the average intensity of these binary patches is more than 0.1 for 10\%-resolution images, or 0.2 for 25\%-resolution images, the patches are considered to be discriminative. We then extract the original patches at that location, and use them for training. Thus, 950,000 patch-pairs are collected to train the network.

\subsection{Validation Data Generation}
In most image registration research, performance is evaluated by measuring the overlap between the registered and target images using the Dice similarity metric. The overlap measurement requires manually segmented ground-truth data. In our case, no ground-truth results are available and manually producing ground-truth for high-resolution CUBIC data is costly and time consuming, and outside the scope of this work. To validate the registration performance in the absence of ground truth, a validation experiment is designed. The test brains are deformed by applying a random Gaussian deformation. All the registration tools are applied on these deformed brains as the source image to register them back to their non-deformed selves.

\section{Method}
\subsection{Dense Deformation Network}
\begin{figure*}
    \centering
          \begin{minipage}[t]{18cm}\includegraphics[width=\linewidth]{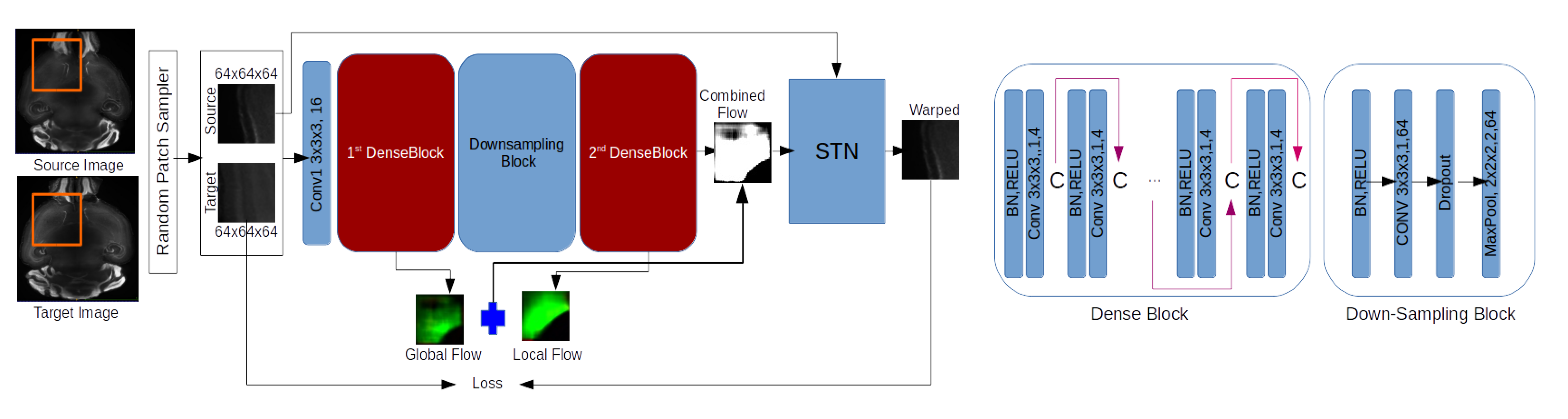}\end{minipage}
    \centering \caption{Architecture of the proposed Dense Deformation Network}
    \label{fig:architecture}
\end{figure*}

In this paper we propose a patch-based deep-learning architecture that is trainable in an unsupervised manner and is scalable to high-resolution images. The proposed network has dense connection blocks and generates deformation fields in two stages unlike other deep-learning based registration methods, most of which employ U-net architectures for deformation generation. In the context of tissue clearing and LSFM, images have discrete intensity distribution with mixed foreground and background. Deformations in these 3D images are small and vulnerable to be affected by the gradual down-sampling in the encoding path of the U-net architecture.
This type of architecture is also used in patch -based segmentation~\cite{Yu2017}. In our application of this architecture, two different levels of deformation are generated and concatenated to obtain the final deformation. The Spatial Transformer Network (STN) is attached to warp the moving image patches at the end of the architecture.
As stated before, the network consists of two dense-blocks as shown in Figure~\ref{fig:architecture}. Each of these dense blocks consists of a number of units. The units are densely connected with each other and each of these units consists of a LeakyReLU activated batch-normalization layer, and a convolution layer. The network takes concatenated source and target image patches as input and passes them through the first dense block without down-sampling. Then a transformation block is used to down-sample the features. A deformation field is generated from the first dense block and is defined as \emph{global flow} since it contains more contextual information. After the transformation block, another dense-block is employed from which another deformation field is generated from these more refined feature banks; we refer to this as the \emph{local flow}. At the end of the network, both of these flows are concatenated to obtain the final deformation field in the x,y and z directions. The source image is then warped with the combined flow by a dense STN. 

\subsection{Loss function}
Similar to~\cite{Balakrishnan2018}, the network is trained with cross-correlation loss and is regularised with diffusion regularisation:

\begin{equation}
    \mathcal{L}_{Loss}(src,tgt,\phi) =\mathcal{L}_{similarity}+\lambda \mathcal{L}_{smooth,}
\end{equation}
where,
\begin{equation}
    \mathcal{L}_{similarity} = -CC(src \circ \phi, tgt),
\end{equation}
and 
\begin{equation}
     \mathcal{L}_{smooth}(\phi) = \sum_{n} \nabla (\phi(p)),
\end{equation}
Here, the diffusion regularisation penalises unrealistic displacements by penalising the x, y and z components of the displacement field. The produced displacement is thus smooth and isotropic.    

\section{ Experiments}
\subsection{Evaluation Metrics}
Most image registration methods are evaluated based on Dice similarity, which requires hand annotated segmentation data. As we explained earlier, this is not currently feasible for tissue-cleared data, and we need an evaluation metric that does not depend on manual annotation, but remains reliable. Cross-correlation and mutual information are therefore selected as evaluation metrics. To further validate the registration accuracy, another experiment is performed where a known deformation is applied on test images and each deformed image is then registered to their non-deformed selves. In addition to this quantitative analysis, we also qualitatively assess the images produced by each method.

\subsection{Competing Methods}
We compare our proposed DDN method against four traditional optimisation-based methods and one learning-based method. 
\subsubsection{ANTS} (Advanced Normalization Tools) uses symmetric diffeomorphic normalisation for non-rigid registration~\cite{AVANTS2008}. For our evaluation, we used cross-correlation as similarity measure with 4 different resolution levels and 100 iterations at each level.
\subsubsection{Elastix} is a large library of different components of image registration~\cite{Klein2010a}. In our experiment, we use parameter settings from~\cite{Hammelrath2016} for rigid, affine and deformable registration.
\subsubsection{NiftyReg} is another high-performing registration tool~\cite{Modat2010}. We use the same settings as~\cite{Xu2016}, except for the number of iterations and intensity threshold. 100 iterations are used for non-rigid registration with free-form deformation model, and we set intensity threshold to 500. 
\subsubsection{IRTK} is one of the early image registration tool with a free-form deformation model~\cite{Rueckert1999a}. We used the same settings as~\cite{Xu2016}, except B-spline control points. Due to the small pixel spacing in the CUBIC dataset, the control point spacing is set to 5mm which is the minimum possible value allowed for this method.
\subsubsection{VoxelMorph} is a recently developed deep-learning-based image registration tool~\cite{Balakrishnan2018}. This method uses the U-net architecture for deformation generation. Unlike our proposed DDN method, the original VoxelMorph takes whole images as input, rather than patches. This limits scalability. In our evaluation, we also considered a patch-based VoxelMorph re-design.
  
\subsection{Experimental Setup}
Our proposed DDN network is developed in Keras with a tensorflow backend. The network is trained and tested in a High-Performance Computing environment with 200 hours walltime, 64 GB RAM, 12 GB Video RAM in Tesla K40m GPU and a single core 2.66GHz 64bit Intel Xeon processor.   

\section{Results}
The performance of the network is evaluated on two different resolution scales, 10\% and 25\% respectively. We compared the quantitative performance of the registration methods by normalised cross-correlation and mutual information for the test brains. For testing the methods we selected three brains from the dataset. For convenience, these are referred to as brains ``001'', ``002'' and ``003''. The performance scores showed in all tables are measured by taking brain ``003'' as the target image and the other two brains as the source image.   
To evaluate the qualitative performance, we show the same brain slice extracted from all test brains and overlaid on the target brain with different colour map. In all of our experiments, the target brain is shown in red, and the registered brains are shown in green. A perfect overlap therefore appears in yellow in the overlaid representation. Note that the posterior part of the brains (cerebellum region) varies significantly between brains, and dissimilarity in this region is expected. For the visual assessment we therefore focus on anterior regions such as the hippocampal formation or the striatum. To facilitate the qualitative evaluation process, we consider not only the whole slice but also specific patches from these regions.\\

For all the methods, we extract patches from the selected regions from both registered image and reference image and calculate the difference between these patches. The difference image contains intensity values in the range +1 to -1 (since all brain volumes are intensity normalised from 0 to 1). When the alignment is accurate, the intensity difference should be 0 in that `difference image'. For proper visualisation of the difference image, we transform the intensities into the range 0 to 255 using a linear triangular function centred at zero. Thus, the resulting transformed images are white in accurately aligned regions and dark in misaligned regions.

\subsection{Performance at 10\% Resolution}
\subsubsection{Quantitative results}
Table \ref{Tab:10_quantitative} shows the performance of all methods. The performance varies across the CC and MI scores. In cross-correlation scores and for brain ``001'', traditional methods obtained the best scores: 0.9589 (Elastix) and 0.9494 (ANTS). Our DDN method is in third position with 0.9093. The original VoxelMorph obtained 0.8962 and its patch-based version obtained 0.8107; both are higher than IRTK. In terms of mutual information, ANTS is the best performer with 0.8057. Elastix and NiftyReg obtained second and third respectively with 0.7899 and 0.6806. Our proposed DDN obtained 0.6098 in mutual information score.
For brain ``002'' the results are largely similar. ANTS is the top scorer in terms CC. Elastix and our proposed DDN obtained very similar scores. In MI score proposed DDN obtained 0.6602. The rank of the other tools remains as for brain ``001''.  
In all measures, DDN outperforms both versions of VoxelMorph. In the remainder of the article, unless otherwise specified, we only use the patch-based version of VoxelMorph (as it is the only one that can scale to larger input sizes).\\

	\begin{table*}[t]
		\begin{center}
			\caption{\small  Performance comparison at 10\% Resolution}
			\label{Tab:10_quantitative}
			\small
			\begin{tabular}{l c c c c}
				\hline
				Methods & brain 1 (CC) &brain 1 (MI) & brain 2 (CC) & brain 2 (MI)\\
				\hline
				BEFORE REGISTRATION     &0.8056		&0.3865		    &0.7716		&0.3619\\
				ANTS					&0.9494		&0.8057			&0.9596		&0.8241\\
		        Elastix		    		&0.9589		&0.7899			&0.9388		&0.8104\\
		        NiftyReg	    		&0.8623		&0.6806			&0.8489		&0.6526\\
		        IRTK		    		&0.8013		&0.5340			&0.6958		&0.5050\\
		        VoxelMorph				&0.8962 	&0.3401			&0.9143 	&0.3578\\
		        VoxelMorph (patch-based)       &0.8107		&0.4543 		&0.8405		&0.4960\\
				DenseDeformation (DDN)   &0.9093		&0.6098		    &0.9258		&0.6602\\
				\hline
			\end{tabular}
		\end{center}
	\end{table*}
	
\subsubsection{Validation of performance}
The results of the validation experiment is shown in Table \ref{Tab:10_validation}. In this experiment, we compared the CC and MI scores of the registered-deformed version of each brain to its original self.  ANTS still ranks first in CC scores for brains ``001'' and ``003''. Elastix obtained the best CC score in brain ``002''. Our proposed DDN network obtained 0.8664, 0.7826 and 0.9826 respectively. The VM network had the lowest performance among all tools (0.8261, 0.6987, 0.9388). In terms of MI score, DDN network obtains scores 0.4383, 0.3490 and 0.6005. This is better than VoxelMorph, but not as good as traditional approaches.
The performance of the deep-learning based tools in the validation test is quite low compared to their traditional competitors. The reason behind this is the deformation applied is global in nature, which disadvantages them. The learning-based tools are more suited to capturing local deformations constrained to the small patches used to train the networks. In this context, the important result is therefore that DDN still outperforms VoxelMorph.\\

\begin{table*}[ht]
	\begin{center}
		\caption{\small  Validation of Registration at 10\% Resolution}
		\label{Tab:10_validation}
		\small
		\begin{tabular}{l c c c c c c}
			\hline
			Methods & brain 1 (CC) &brain 1 (MI) & brain 2 (CC) & brain 2 (MI) & brain 3 (CC) & brain 3 (MI)\\
			\hline
			ANTS              &0.9836	&0.6424   &0.9716	&0.6635   &0.9973   &0.7538\\
			Elastix           &0.9757   &0.5880   &0.9731	&0.6062   &0.9956	&0.6955\\
			NiftyReg          &0.9720	&0.6068   &0.9641	&0.6416   &0.9923	&0.6556\\
		           IRTK              &0.9448   &0.6453   &0.9323   &0.6522   &0.9445   &0.6489\\
		    VoxelMorph        &0.8261   &0.3706   &0.6987   &0.2980   &0.9388  &0.4523\\ 
		    DenseDeformation  &0.8664	&0.4383   &0.7826	&0.3490   &0.9826   &0.6005\\
			\hline
		\end{tabular}
	\end{center}
\end{table*}

\subsubsection{Qualitative results}
\label{sec:10_qualitative}
\begin{figure*}
    \centering
    \includegraphics[clip,trim={00mm 10mm 00mm 00mm},width=\linewidth]{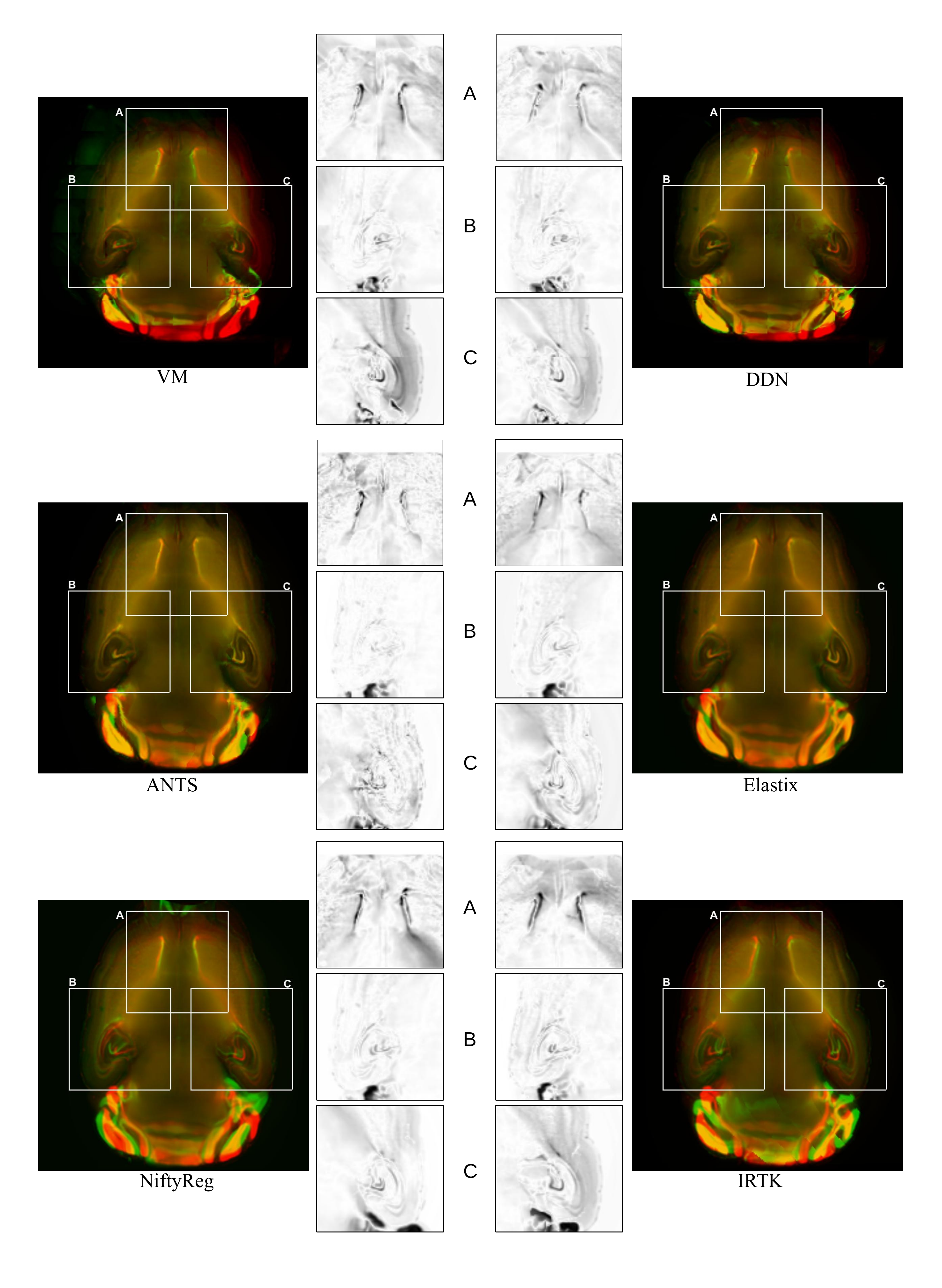}
    \caption{Visual comparison at 10\% resolution}
    \label{fig:10_compareDiff}
\end{figure*}

The qualitative performance of the proposed DDN and five other tools at 10\% resolution is shown in Figure~\ref{fig:10_compareDiff}. The first row shows the results of our proposed architecture and of the patched-based VoxelMorph. The second and third rows contain the results for optimisation-based tools. 
When considering the whole brains, ANTS and Elastix have the best overall performance, while NiftyReg and IRTK are the worst performers. The deep learning based methods are in the middle of the range, with a slight advantage to DDN. This is in line with the quantitative measures.

In patch A (striatum region), the boundary between the striatum ventral region and the lateral septal complex is a clear visual indicator of the alignment quality. ANTS and Elastix perform best. Ordered by increasing misalignment of that boundary, the next tools are DDN, NiftyReg, VoxelMorph, and IRTK.

In patches B and C (left and right hippocampal formations, respectively), Ammon's horn and the dentate gyrus are clear visual features. The alignment is excellent in patch B for ANTS, and very good for Elastix and DDN. VoxelMorph is slightly worse, while NiftyReg and IRTK both struggle. In patch C, the alignment quality is generally worse, apart from ANTS and Elastix.

\subsection{Performance at 25\% Resolution}
\subsubsection{Quantitative results}
\label{sec:25_qualitative}
The registration performance of all tools at 25\% resolution is shown in Table~\ref{Tab:25_quatitative}. 
As before, ANTS and Elastix obtain the best results among classic methods, in terms of both CC and MI. Our proposed DDN method compares favourably. For brain ``001'', the CC performance is close (0.8337, compared to 0.8712 for ANTS and 0.8686 for Elastix). All other methods scored below 0.79. In terms of MI, DDN ranked second overall (1.3583 for ANTS, 1.1765 for DDN, and all other methods below 1.16). For brain ``002'', DDN ranked first for CC (0.9438, against 0.9391 for ANTS and less than 0.92 for all other methods) and second for MI (1.5003, compared to 1.5918 for ANTS, and at most 1.40 for all other methods).
For both test brains, our proposed DDN architecture performed much better than its technically closest rival, VoxelMorph. Overall, the performance measure at 25\% resolution indicates the potential ability of our proposed architecture compared to both traditional and deep-learning based methods.\\

\begin{table*}[ht]
	\begin{center}
		\caption{\small  Performance comparison at 25\% Resolution}
		\label{Tab:25_quatitative}
		\small
		\begin{tabular}{l c c c c}
			\hline
			Methods & brain 1 (CC) &brain 1 (MI) & brain 2 (CC) & brain 2 (MI)\\
			\hline
			BEFORE REGISTRATION     &0.6763  	&0.9147         &0.8016  	&1.2006\\
			ANTS			    	&0.8712 	&1.3583			&0.9391		&1.5918\\
			Elastix				    &0.8686		&1.1542			&0.9102		&1.2903\\
			NiftyReg			    &0.7166		&1.1319			&0.7570 	&1.1677\\
			IRTK				    &0.6148		&0.8205			&0.6713		&0.9154\\
		    VoxelMorph          &0.7877		&1.0620			&0.9126		&1.4009\\
		    DenseDeformation         &0.8337		&1.1765			&0.9438		&1.5003\\
			\hline
		\end{tabular}
	\end{center}
\end{table*}

\subsubsection{Validation of Performance}
The validation results at 25\% resolution are shown in Table~\ref{Tab:25_validation}. The results are largely similar to those obtained at 10\% resolution.
ANTS remains the top performer among classic methods, with NiftyReg and Elastix also providing good results. Our proposed DDN network obtains comparable results for CC, and slightly lower for MI. It outperforms both IRTK and VoxelMorph on all brains and both metrics. Given the nature of the deformation, the main result is, again, that DDN provides better results than VoxelMorph.\\

\begin{table*}[ht]
\begin{center}
 \caption{\small Validation  of Registration at 25\% Resolution}
 \label{Tab:25_validation}
 \small
 \begin{tabular}{l c c c c c c}
  \hline
             Methods & brain 1 (CC) &brain 1 (MI) & brain 2 (CC) & brain 2 (MI) & brain 3 (CC) & brain 3 (MI)\\
             \hline
             ANTS              &0.9965  &2.0971 &0.9986  &2.3362  &0.9922  &2.1848\\
             Elastix           &0.9859  &1.7062 &0.9958  &1.9615  &0.9821  &1.8004\\
             NiftyReg          &0.9987  &2.0148 &0.9995  &2.1910  &0.9890  &2.1985\\
             IRTK              &0.8024   &1.2238 &0.9383  &1.8733  &0.7238  &1.0947\\
             VoxelMorph        &0.9661   &1.6388    &0.9687  &1.6385    &0.9639  &1.7660\\
             DenseDeformation  &0.9767   &1.6733    &0.9888  &1.9153  &0.9772  &1.8158\\
             \hline
 \end{tabular}
\end{center}
\end{table*}

\subsubsection{Qualitative results}
\begin{figure*}
    \centering
    \includegraphics[clip,trim={00mm 10mm 00mm 00mm},width=\linewidth]{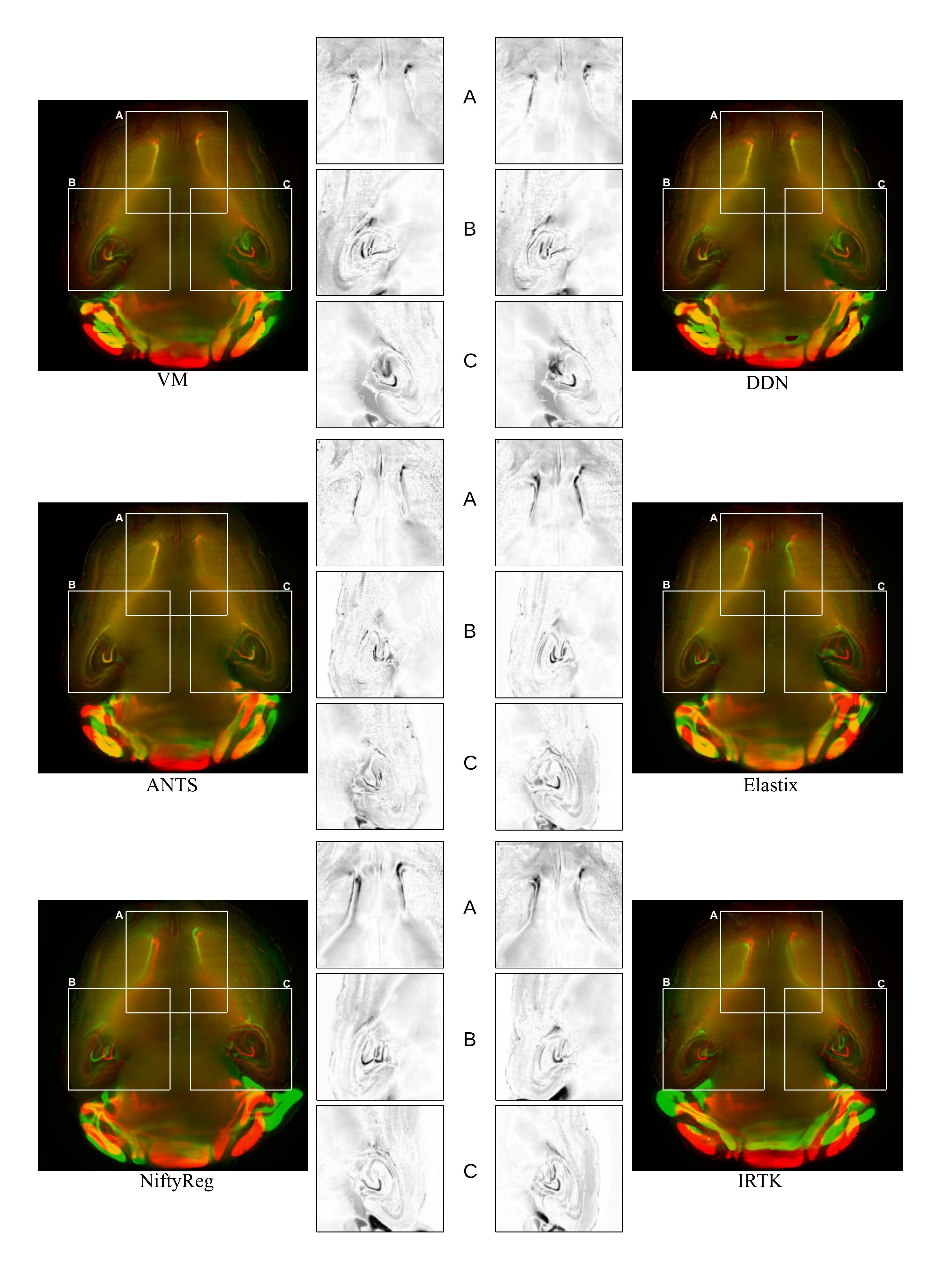}
    \caption{Visual comparison at 25\% resolution}
    \label{fig:25_compareDiff}
\end{figure*}

The qualitative results at 25\% resolution are shown in Figure~\ref{fig:25_compareDiff}. The first row shows the results of our proposed architecture and of the patched-based VoxelMorph. The second and third rows contain the results for optimisation-based tools. Overall, these results confirm those observed at 10\% resolution.

When considering the whole brains, ANTS continues to have the best overall performance, while NiftyReg and IRTK are still the worst performers. Elastix and the two methods based on deep learning are in the middle of the range. DDN and Elastix have similar level of misalignment (but in different regions), with VoxelMorph having more errors than both. This is consistent with the quantitative measures.

In patch A (striatum region) we can still use the boundary between the striatum ventral region and the lateral septal complex as a visual marker of the alignment quality. ANTS and DDN perform best. Ordered by increasing misalignment of that boundary, the next tools are VoxelMorph, Elastix, IRTK and NiftyReg.

In patches B and C (left and right hippocampal formations, respectively), we still focus on Ammon's horn and the dentate gyrus. The alignment is good in patch B for ANTS, and fair for Elastix and DDN. All other tools have a poor performance. In patch C, the alignment quality is generally worse, with only ANTS producing acceptable results. High-resolution registration remains a challenge for all tools.

As for previous tests, our proposed DDN method has better performance than the other deep-learning based method (VoxelMorph) and several classic approaches (NiftyReg, IRTK, even Elastix at times), but does not match ANTS.

\section{Efficiency}
The computational performance of our proposed DDN and other tools are shown in Table \ref{Perform_in_time} for both 10\% and 25\% tests. While ANTS produces the best results both quantitatively and qualitatively, this comes at a cost: it is one of the slowest tools. At 10\% resolution it takes about 39 minutes, while for 25\% resolution it needs almost 9 hours per brain, which would not be practical for studies involving a large number of brains. IRTK and NiftyReg are also slow, but were already not a promising option after the quality assessment. Elastix is not as good as ANTS, but it is faster: approximately 2 minutes and 40 minutes at 10\% and 25\% resolution, respectively. This still represents an important bottleneck at 25\% resolution for large studies, and would not scale to higher resolutions.

Compared to the optimisation-based tools, both methods using deep learning are significantly more efficient. This is normal as, once the network is trained, a new registration is simply a forward pass of matrix multiplication. Our DDN architecture has more convolutional layers than VoxelMorph, so it is slightly slower (12 seconds and 1.5 minute, compared to 5 and 55 seconds), but it produces better results both quantitatively and qualitatively.
The computational performance of both methods depends not only on the size of the network architecture but also on the number of patches extracted from the source and target images. In our experiments, we used a 50\% overlap for both networks. Reducing the overlap would produce faster, but degraded, results.

\begin{table}[h!]
	\small
	\begin{center}
		\caption{Registration Performance in Time}
		\label{Perform_in_time}
		\begin{tabular}{l c c}
			\hline
			Methods & 10\%  & 25\%\\
			\hline
			ANTS		&00:39:26  	&08:32:24\\
			Elastix 	&00:02:09  	&00:40:17\\ 
			NiftyReg	&00:08:46   &02:10:45\\
			IRTK		&00:05:11   &11:29:16 \\
			VM          &00:00:05   &00:00:55\\
			DDN         &00:00:12   &00:01:35\\
			\hline
		\end{tabular}
	\end{center}
\end{table}

\section{Conclusion}
In this paper we proposed a patch-based 3D registration framework that consists of densely connected convolutional layers. The proposed architecture utilises two steps of deformation generation and produces a dense deformation field. The unsupervised patch-based training strategy gives the network the ability to learn displacement without requiring any ground truth data. 
The network performs well in all quantitative tests, and better as the resolution increases. This indicates a better ability to handle high-resolution data typical of samples obtained from tissue clearing. The qualitative results at two different resolution scales confirm that the proposed DDN network can better handle the complexity of tissue-cleared data than the existing VoxelMorph architecture.
While some traditional tools (most notably ANTS) can produce higher-quality registration, the ability of the DDN network to register images with comparable accuracy in minutes rather than hours makes it a valuable tool for practical applications.

Further research is needed in deep learning based image registration methods. DDN and VoxelMorph both struggle with large deformations. Investigation of loss functions and regularisation parameters is a possible direction. Another limitation is that the patch-based training can also generate discontinuous deformation fields, which may lead to unnatural images. Investigation of patch-by-patch continuous deformation generation will also be considered.    

\section*{Acknowledgment}
Computational resources and services used in this work were provided by the HPC and Research Support Group, Queensland University of Technology, Brisbane, Australia.
\ifCLASSOPTIONcaptionsoff
  \newpage
\fi

\bibliographystyle{ieeeconf}
%
\bibliography{new_ref_v2.bib}

%
%
%

\end{document}